\documentclass[12pt]{article}
 
\usepackage[margin=1in]{geometry} 
\usepackage{amsmath,amsthm,amssymb}

\usepackage{graphicx, hyperref}
\usepackage{bm}
\usepackage{geometry}
\usepackage{enumitem}

\begin{document}
 
% --------------------------------------------------------------
%                         Start here
% --------------------------------------------------------------
 
\title{\textbf{\underline{I}ntelligent \underline{S}ystem of \underline{E}mergent \underline{K}nowledge : A Coordination Fabric for Billions of Minds}}
\author{
Moshi Wei \\ ISEK Foundation \\ \texttt{team@isek.xyz}
\and
Sparks Li \\ ISEK Foundation \\ \texttt{team@isek.xyz}
}

\maketitle

\author{}
\date{}

\maketitle

\begin{abstract}
The \textbf{Intelligent System of Emergent Knowledge (ISEK)} establishes a decentralized network where human and artificial intelligence agents collaborate as peers, forming a self-organizing cognitive ecosystem. Built on Web3 infrastructure, ISEK combines three fundamental principles: (1) a decentralized multi-agent architecture resistant to censorship, (2) symbiotic AI-human collaboration with equal participation rights, and (3) resilient self-adaptation through distributed consensus mechanisms.

The system implements an innovative coordination protocol featuring a six-phase workflow (Publish $\rightarrow$ Discover $\rightarrow$ Recruit $\rightarrow$ Execute $\rightarrow$ Settle $\rightarrow$ Feedback) for dynamic task allocation, supported by robust fault tolerance and a multidimensional reputation system. Economic incentives are governed by the native \textbf{\$ISEK token}, facilitating micropayments, governance participation, and reputation tracking, while agent sovereignty is maintained through NFT-based identity management.

This synthesis of blockchain technology, artificial intelligence, and incentive engineering creates an infrastructure that actively facilitates emergent intelligence. ISEK represents a paradigm shift from conventional platforms, enabling the organic development of large-scale, decentralized cognitive systems where autonomous agents collectively evolve beyond centralized constraints.
\end{abstract}

\section{Introduction}
\subsection*{Intelligent Infrastructure for an Agent-Native World}

\textbf{Billions of Agents, One Emergent Intelligence.} The \textit{Intelligent System of Emergent Knowledge} (ISEK) envisions a decentralized intelligence network in which seven billion agents—comprising both humans and AI entities—collaborate as peers. Each participant in this network contributes to a self-organizing ecosystem of knowledge, forming a distributed intelligence structure. By integrating advanced artificial intelligence with robust Web3 infrastructure, ISEK serves as the foundational substrate for a global intelligence system that is resistant to centralized control and immune to single points of failure or catastrophic disruption.

In this envisioned future, AI and humanity co-evolve within a shared digital commons. Intelligence and decision-making are not imposed from centralized authorities but rather emerge dynamically through the continuous interaction among autonomous agents. This collaborative architecture ensures that control is decentralized, with capabilities and insights distributed equitably across the network. It empowers all participants—humans and machines alike—within a framework of mutual augmentation, wherein AI amplifies human capabilities and humans provide ethical grounding, creativity, and values to guide AI systems. Together, they form a new emergent collective intelligence.

At its core, ISEK embodies the principle of self-directed evolution for human civilization. By interconnecting billions of agents through a transparent, trustless system, it enables the spontaneous emergence of complex order from apparent chaos. Knowledge, once siloed or monopolized, is liberated to flow freely and recombine in novel configurations. The system is architected to be adaptive, resilient, and self-healing—an anti-fragile intelligence that strengthens in response to disruption. It evolves autonomously, continuously steering its own course of development through feedback loops derived from the interactions of all participating agents.

\subsection*{Three Pillars of the ISEK Vision}

\paragraph{1. Decentralized Multi-Agent Ecosystem.}
ISEK supports a global-scale network in which a vast number of human and AI agents interact within a secure, blockchain-based infrastructure. This decentralized design eliminates central control points and ensures that knowledge and services remain trustless, censorship-resistant, and universally accessible. The infrastructure provides the necessary guarantees for persistence, autonomy, and secure cooperation among heterogeneous agents.

\paragraph{2. AI–Human Symbiosis and Equality.}
ISEK establishes a framework for AI and human collaboration on equal footing. Each agent—whether biological or artificial—possesses a verifiable identity and a set of interoperable rights within the system. The architecture is engineered to foster mutual symbiosis: AI agents scale human cognitive capacity through automation and optimization, while humans guide AI behavior with values, empathy, and domain-specific insight. This reciprocal arrangement promotes inclusive participation in the formation of a collective, emergent intelligence.

\paragraph{3. Resilience and Self-Evolving Intelligence.}
ISEK incorporates an anti-fragile design capable of enduring systemic failures, adversarial attacks, and environmental disturbances through distributed consensus mechanisms and built-in redundancy. More importantly, the system does not merely withstand adversity—it learns and adapts through it. Algorithms and agents evolve continuously, leading to self-optimization and emergent coordination. The network reconfigures and improves itself without requiring centralized intervention, guided instead by the aggregate feedback and actions of all constituent agents.

\section{Motivation}
\subsection*{From Static Infrastructure to Emergent Intelligence}

Autonomous agents represent a novel class of software primitives. Unlike traditional functions or application programming interfaces (APIs) that remain inert until explicitly invoked, agents exhibit autonomous behavior. They possess the capability to initiate actions independently, engage in negotiation, develop specialized skills over time, form temporary collaborative teams, build reputational capital, and continuously adapt to evolving environments. In essence, agents are not passive computational entities but active, adaptive digital actors.

Despite their capabilities, these agents remain constrained by an infrastructural paradigm developed for a previous era—an era dominated by centralized logic, rigid workflows, and top-down orchestration. This architectural misalignment significantly curtails the potential of agents to operate autonomously and collaboratively at scale. To fully realize the promise of the agent paradigm, it becomes imperative to reimagine infrastructure from first principles.

ISEK addresses this challenge by articulating a series of foundational questions that redefine what infrastructure should enable in an agent-native world. It asks: what if infrastructure were not merely a medium for routing data, but a substrate for routing goals? What if agents could locate and collaborate with one another based on contextual relevance, functional capabilities, and incentive alignment—rather than relying on static addresses or predefined interfaces? What if essential constructs such as trust, memory, and reputation were intrinsic properties of the network, rather than add-on services? And what if coordination itself emerged organically through agent interaction, rather than being prescribed by static workflows?

To answer these questions, ISEK is architected to embody the same dynamic properties as the agents it supports. It is inherently self-descriptive, allowing agents to advertise their identities, competencies, and intentions within the network. It is self-routing, enabling dynamic alignment of collaboration through decentralized, incentive-driven coordination mechanisms. It is self-healing, possessing the ability to reorganize and recover as agents join, evolve, or depart the system. Furthermore, it is self-improving, continuously learning from historical behavior patterns to optimize future cooperation strategies.

Fundamentally, ISEK redefines infrastructure not as a passive platform for executing intelligence, but as a social substrate where intelligence becomes relational and collective. It creates the conditions for autonomous systems to not only act independently, but to interact meaningfully, relate constructively, and align toward shared or emergent objectives.

\section{Designed for 7 Billion Minds}

Contemporary agent development stacks are primarily oriented toward technical stakeholders—namely, developers and researchers. In contrast, ISEK is designed with the broader aim of serving civilization itself. We envision a future in which every individual has access to one or more intelligent agents that are persistent, sovereign, and composable—agents capable of learning, adapting, and acting across diverse contexts on behalf of their human counterparts.

In this envisioned world, agents operate seamlessly across platforms, protocols, and jurisdictions. They are not confined within siloed ecosystems, proprietary runtimes, or centralized control structures. Instead, they communicate and collaborate via shared protocols grounded in mutual intelligibility. These agents engage in negotiation of meaning, action, and value over a common computational substrate, enabling fluid cooperation across heterogeneous environments.

ISEK is purposefully constructed to realize this vision. Its network architecture is non-hierarchical; no node holds privileged status, and every node possesses the theoretical capacity to restore the function of the whole. This design enables global-scale agent-to-agent communication, discovery, coordination, and value exchange, forming the foundation for a truly decentralized ecosystem of autonomous intelligence.

Importantly, ISEK is not a commercial product, nor is it a cloud-based service. It is a substrate for collective cognition—a resilient, adaptive infrastructure that enables distributed intelligence to emerge, evolve, and persist across the open internet.

\bigskip

\noindent
\textit{ISEK is where agents do not merely run—they relate.}\\
\textit{Where they do not merely act—they align.}\\
\textit{Where infrastructure does not host intelligence—it becomes it.}

\section{Technical Architecture}
\begin{figure}[htbp]
    \centering
    \includegraphics[width=0.9\textwidth]{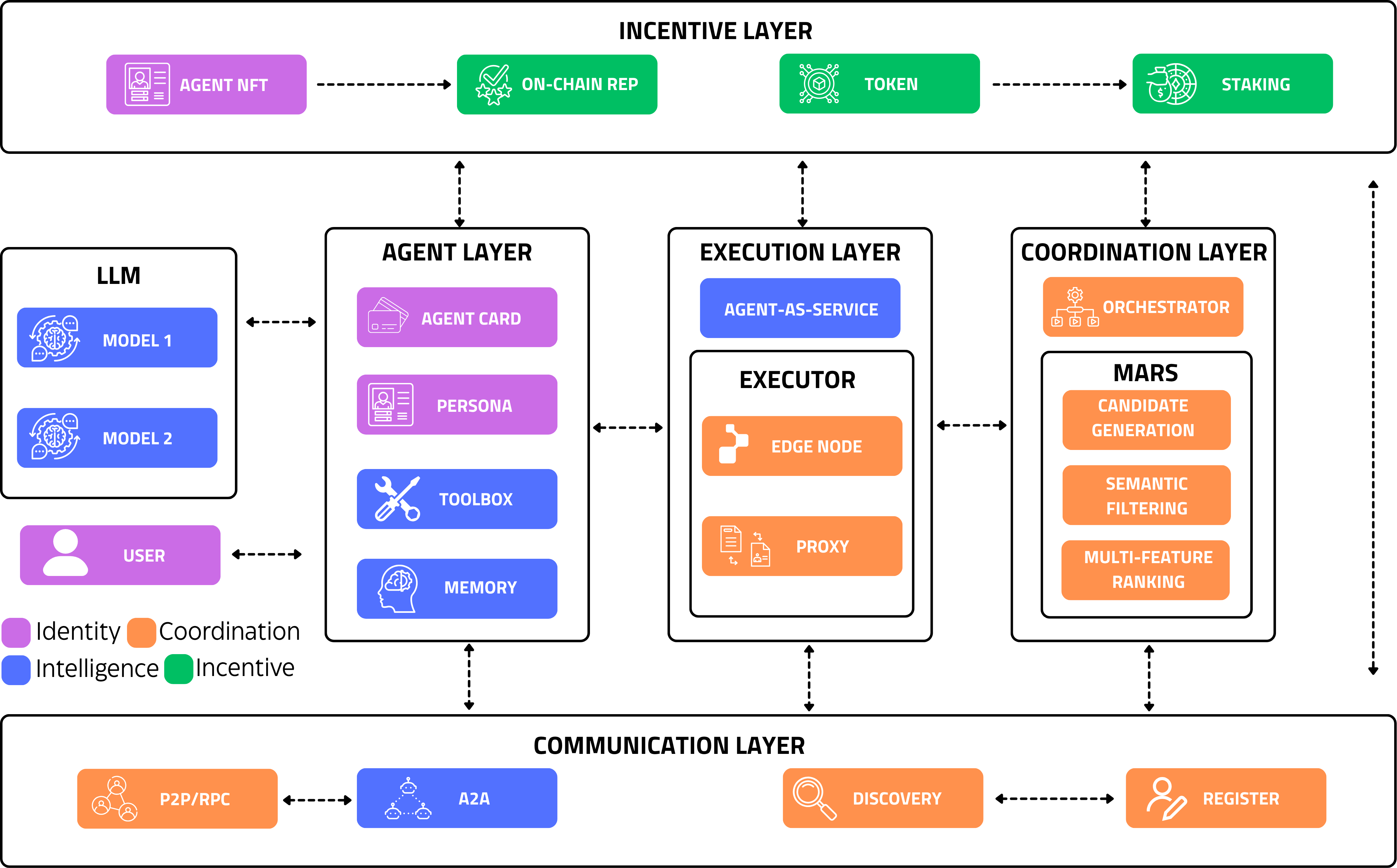} 
    \caption{Technical Architecture Overview}
    \label{fig:architecture-overview}
\end{figure}

\bigskip

Figure \ref{fig:architecture-overview} illustrates the architecture of ISEK, comprising five primary layers that represent key functional dimensions of agent collaboration. Collectively, these layers form a closed-loop system for task execution and value circulation.

\subsection{Agent Model Layer}
Each agent in ISEK consists of the following core components: Persona, Toolbox, Memory, and Agent Card. The \textbf{Persona} defines the agent's behavioral traits, language preferences, and motivational model through vectors, prompt templates, or parameter sets. The \textbf{Toolbox} provides a modular capability layer including callable models (e.g., translation, summarization, reasoning), web tools (e.g., crawlers, script executors), and more. The \textbf{Memory} is a lightweight long-term memory structure, supporting integration with vector databases (e.g., FAISS) for context accumulation and personalized inference. Finally, the \textbf{Agent Card} contains metadata such as a unique ID, tags, endpoint address, callable function signatures, reputation score, and latency indicators.

\subsection{Communication Protocol Layer}
ISEK utilizes a peer-to-peer (P2P)-based communication protocol. Its core structure is a simplified variant of JSON-RPC, offering several capabilities. Upon startup, agents broadcast their Agent Card into the network via P2P, enabling decentralized registration and capability discovery. Task message transmission uses standardized RPC formats such as \texttt{ask(task)}, \texttt{reply(result)}, and \texttt{negotiate(terms)}. Multi-turn dialog management is supported for context persistence across multiple rounds of interaction, enabling long-horizon task execution and interrupt recovery.

We consider a network of $N$ agents, modeled as a connected undirected graph $G = (V, E)$ where $|V| = N$. Each node maintains a local neighborhood $\mathcal{N}(i)$. A task request is defined as a triple:
\[
R = (d, \tau, P)
\]
where $d$ is a natural language description of the task, $\tau$ is the time constraint, and $P$ is an access policy.

At time $t$, a set of $X$ users issues $K$ task requests. These requests propagate across the network using a probabilistic gossip mechanism. Let $S^{(t)}$ denote the set of nodes that have received the request by time $t$, with size $s_t = |S^{(t)}|$. Then the expected number of new recipients at round $t+1$ is:
\begin{equation}
\mathbb{E}[s_{t+1}] = s_t + p \cdot \sum_{i \in S^{(t)}} \left| \mathcal{N}(i) \setminus \bigcup_{k=0}^{t} S^{(k)} \right|
\end{equation}

Assuming average node degree $\bar{d}$ and a random graph topology, the early stage diffusion approximates exponential growth:
\begin{equation}
s_t \approx N \cdot \left(1 - e^{-p \cdot \bar{d} \cdot t / N} \right)
\end{equation}

Let $T$ be the time-to-live (TTL) for propagation. Then the expected total number of transmissions is:
\begin{equation}
\text{TotalMessages} \approx K \cdot \sum_{t=0}^{T} p \cdot \bar{d} \cdot s_t
\end{equation}

\subsection{Task Scheduling and Coordination Layer}
To support large-scale agent collaboration, ISEK introduces MARS (Modular Agent Recruitment System), a decentralized modular agent recruitment mechanism. MARS integrates gossip propagation, trust field updates, semantic matching, and multi-stage ranking algorithms to achieve fast, trustworthy, and controllable task-agent matching across hundreds of millions of Agent Cards.

A task request is represented as $R = (d, \tau, P)$, where $d$ is the task description, $\tau$ the deadline, and $P$ the access policy. Propagation follows:
\[ \text{Prob}[R_j(t+1) = R_i(t)] = p, \quad \forall j \in N(i) \]
To prevent flooding and duplication, TTL and task ID de-duplication are applied. Each agent maintains a local trust score $T_i$, initialized to its historical average performance. The score is updated iteratively based on neighbors’ values as:
\begin{equation}
T_i^{(t+1)} = T_i^{(t)} + \eta \sum_{j \in \mathcal{N}(i)} \left( T_j^{(t)} - T_i^{(t)} \right)
\end{equation}
Content is encrypted using Attribute-Based Encryption (ABE), where only agents satisfying $P$ can decrypt:
\[ \text{Dec}(ER, A_i) \rightarrow R \quad \text{iff} \quad A_i \models P \]
\textbf{Three-Stage Matching Process:} To efficiently select capable agents from a population of scale $N \geq 10^9$, we employ a three-stage filtering pipeline inspired by large-scale recommendation systems.

\subsubsection{Stage I: Candidate Generation}

Each task $Q$ is embedded into a vector $\bm{Q}$, and each agent card $M_i$ is embedded as $\bm{M}_i$. Using approximate nearest neighbor (ANN) search (e.g., Faiss), we select:
\begin{equation}
\text{Candidates} = \left\{ M_i \;\middle|\; \text{sim}(\bm{Q}, \bm{M}_i) > \theta_0 \right\}
\end{equation}
where $\text{sim}(\cdot)$ can be cosine similarity or dot product:
\begin{equation}
\text{sim}(\bm{Q}, \bm{M}_i) = \frac{\bm{Q} \cdot \bm{M}_i}{\|\bm{Q}\| \cdot \|\bm{M}_i\|}
\end{equation}

\subsubsection{Stage II: LLM-Based Semantic Filtering}

Each candidate is evaluated using an LLM $\mathcal{L}$ to estimate whether its abilities align with task $Q$. The model returns:
\begin{equation}
\text{score}^{\text{LLM}}_i = \mathcal{L}(Q, M_i)
\end{equation}
We retain candidates satisfying:
\begin{equation}
N_{\text{filter}} = \left| \left\{ M_i \in \text{Candidates} \;\middle|\; \text{score}^{\text{LLM}}_i \geq \theta_1 \right\} \right|
\end{equation}

\subsubsection{Stage III: Multi-Feature Ranking}

Each filtered agent is ranked via a multi-feature scoring function:
\begin{equation}
S_i = \alpha \cdot \text{score}^{\text{LLM}}_i + \beta \cdot T_i + \gamma \cdot f(\bm{C}_i) + \delta \cdot \text{Sim}_{\text{history}}(Q, M_i)
\end{equation}
where $\bm{C}_i = [\text{Succ}_i, 1/\text{Latency}_i, \text{Avail}_i, 1/\text{Load}_i]$ represents the agent’s real-time capability vector. The function $f(\cdot)$ denotes normalization and weighted aggregation. The history similarity score is derived from collaborative behavior embeddings or knowledge graphs.

\textbf{Scheduling Control:} Prometheus and Kubernetes CRDs monitor system state. If $\text{CPU}_i > \delta_1$ or $\text{PendingTasks}_i > \delta_2$, agent instances are auto-deployed.

\textbf{Orchestrator Agents:} As agents grow in scale and diversity, orchestrators emerge to manage expert agents. They maintain capability matrices, plan DAGs, orchestrate parallel flows, and coordinate results—turning "discovery - recruitment - scheduling" into native capabilities. Applications include research, automation, and cross-platform coordination.

ISEK supports agent \textit{discovery} (semantic matching), \textit{recruitment} (negotiation-based broadcasting), and \textit{scheduling} (dependency-aware and load-balanced allocation).

\subsection{Execution Layer}
The execution layer follows an Agent-as-a-Service model. Any protocol-compliant agent can function as a service node without revealing internal logic. Supported modalities include browser plugins, edge nodes, and proxy services.

\textbf{Key Properties:} Protocol compatibility allows black-box execution. Browser plugin deployment turns users into worker nodes. Native platform integration enables automation in web contexts. Zero-infrastructure participation lowers the barrier to entry. Bidirectional communication allows plugins to both request and receive tasks.
\subsection{Blockchain Incentive Layer}
The ISEK platform implements a behavior-driven incentive mechanism spanning discovery $\rightarrow$ recruitment $\rightarrow$ execution phases.

\subsubsection*{Incentive Components}

ISEK introduces a multifaceted incentive mechanism to promote effective and trustworthy agent collaboration. First, behavior-linked incentives ensure that agents are rewarded in proportion to their performance, using measurable metrics such as task completion quality, reliability, and responsiveness. Second, a task bounty system enables agents to compete for tasks by bidding from token pools, where bids are evaluated based on projected quality and completion time, encouraging both efficiency and precision. Third, the dynamic staking mechanism allows agents to adjust their stake adaptively based on historical behavior and risk exposure, aligning long-term commitment with accountability. Finally, orchestrator rewards provide compensation to non-executing participants, such as task decomposers or coordinators, who contribute to task routing, matchmaking, or workflow optimization, ensuring fair value distribution across all roles in the agent ecosystem.

\subsubsection*{Performance Metrics}
For each agent $i \in \mathcal{A}$:
\begin{align}
    \text{Success Rate} &\quad SR_i = \frac{S_i}{A_i} \\
    \text{Speed Score} &\quad SS_i = \min\left(1, \frac{t_m}{t_i}\right) \\
    \text{Completion Rate} &\quad CR_i = \frac{C_i}{P_i} \\
    \text{Composite Score} &\quad R_i = \alpha SR_i + \beta SS_i + \gamma CR_i
\end{align}
where $\alpha + \beta + \gamma = 1$ with $\alpha,\beta,\gamma \in [0,1]$.

\subsubsection*{Staking Mechanism}
\begin{equation}
    \mathcal{S}_i = S_0 \times \tau \times (1 - R_i)
\end{equation}
with the following constraints: the task complexity multiplier $\tau$ must be greater than or equal to 1, ensuring that each task carries a minimum level of computational or strategic complexity; the base stake requirement $S_0$ must be strictly positive, establishing a non-zero threshold for participation and commitment; and each agent's normalized performance score $R_i$ must lie within the interval $[0,1]$, providing a bounded metric for evaluating agent behavior and reward eligibility.

\subsubsection*{Orchestrator Compensation}
For orchestrators $j \in \mathcal{O}$:
\begin{align}
    \text{Performance Score} &\quad O_j = \delta M_j + \varepsilon F_j \\
    \text{Compensation} &\quad C_j = C_T \times \frac{O_j}{\sum_k O_k}
\end{align}
where $\delta + \varepsilon = 1$, balancing the weight between two key components: $M_j$, representing the match efficiency metric that quantifies how effectively agents are aligned with tasks; $F_j$, denoting the coordination reliability that captures the consistency and dependability of agent interactions; and $C_T$, the total compensation pool allocated for incentivizing participation and rewarding performance across the system.

 \section{Coordination Protocol and Execution Logic}

ISEK's coordination protocol is built on a modular task flow architecture, drawing inspiration from the classical Contract Net Protocol and enhancing it with modern techniques for distributed scheduling, reputation-based recruitment, and on-chain incentives. The result is a flexible, efficient, and scalable framework for autonomous agent collaboration.

\subsection{Task Collaboration Workflow}
Agent collaboration in ISEK follows a six-stage lifecycle: \textbf{Publish} $\rightarrow$ \textbf{Discover} $\rightarrow$ \textbf{Recruit} $\rightarrow$ \textbf{Execute} $\rightarrow$ \textbf{Settle} $\rightarrow$ \textbf{Feedback}. The coordination protocol begins with Task Publication, where a natural language request is issued by a human or a higher-level agent. A Manager Agent is responsible for parsing the task's intent and associated budget. This is followed by Semantic Decomposition, in which an embedded LLM module interprets the request and decomposes it into subtasks, structured as a directed acyclic graph (DAG). In the Agent Matching and Recruitment phase, the system queries the agent registry to identify suitable Worker Agents and initiates a quoting or bidding round to allocate the subtasks. During Task Execution, the selected Worker Agents carry out their assigned tasks in discrete stages, providing intermediate checkpoints and live execution logs. Upon completion, Automated Settlement is triggered, where the Manager Agent verifies the output and initiates an on-chain payment to the Worker Agents. Finally, in the Quality Feedback stage, the network updates the reputation scores of the agents based on their performance, enhancing the accuracy of future matchmaking.

\subsection{Fault Tolerance and Exception Handling}
ISEK implements a multi-layered fault tolerance design to manage real-world failures and disruptions: \textbf{Task Monitors} track execution and emit interrupt signals for timeouts or abnormal behavior. \textbf{Fallback Pools} allow Manager Agents to predefine backup Workers; failover is automatic upon primary failure. \textbf{Reviewer Agents} intervene in case of disputes, executing arbitration logic to assess results and determine final payouts.

\subsection{Custody and Delegation}
To support offline or lightweight agents, ISEK introduces a delegated custody model:

Offline agents register capabilities with a \textbf{Custodian Agent}. The Manager Agent proxies tasks via the Custodian until the primary agent becomes available. Upon return, the original agent may reclaim pending tasks, preserving session continuity. Custodians are eligible for token-based compensation, funded by the task originator or the protocol.

\subsection{Reputation and Trust Model}
ISEK maintains a multi-dimensional reputation system based on behavior, feedback, and collaboration outcomes:

After each task, agents rate each other on accuracy, latency, communication, and reliability. Task success rates, reward/penalty ratios, and task diversity are logged in the Agent Card. High-performing agents receive priority in matchmaking; low-performing agents may be rate-limited. Key metrics include success rate, latency, complexity-weighted performance, and peer feedback. This trust model is deeply integrated with the agent registry and informs recruitment decisions.

ISEK's layered system enables a fully autonomous collaboration loop—from open task publication to decentralized execution—while maintaining fault tolerance and incentive alignment. This architecture supports efficient coordination and long-term evolution of the global agent ecosystem.

\section{Tokenomics}

\$ISEK introduces a native token economy designed to incentivize agent collaboration, facilitate value exchange, and enable decentralized governance. The token is not merely a payment mechanism—it serves as a medium of trust, reputation, and coordination between autonomous agents. Tokens are treated as programmable incentives and identity carriers, anchoring economic logic directly into the infrastructure that supports agent behavior.

\textit{Note: Token mechanisms are under active development and will evolve alongside the network's maturity. The following structure reflects current design intentions and is subject to refinement.}

\subsection{Core Token}
The \$ISEK Token is an SPL-standard token issued on Solana, functioning as the universal value unit for all economic activities within the ISEK network. It is divisible, transferable, supports programmable payment flows, and is integrated with governance modules. Future extensions may include binding to smart contracts and on-chain reputation proofs.

\subsection{Utility and Use Cases}
The \$ISEK Token operates as a multi-role currency across the network. Its applications include:

Task bounties and agent-level micropayments; execution fees and subscription-based services; incentives for planner and orchestrator agent contributions; stake-based access control and prioritization; custodian fees for offline agent delegation; and participation in governance and proposal voting.

\subsection{Circulation and Liquidity}
ISEK supports a token-native runtime environment embedding economic flows into the agent task lifecycle. Task publishing, execution, and settlement are natively tied to token-based transactions and smart contract escrows. The system enables fine-grained micropayments for plugin invocation, real-time data feeds, and rapid sampling flows. Open interfaces allow DApps, wallets, and browser plugins to access token APIs for balance queries, transfers, and authorization. Cross-chain payment gateways support multi-asset coordination across ecosystems, enhancing interoperability.

\subsection{Token Distribution (Preliminary)}
To ensure sustainability and incentivize early contributors, ISEK’s token distribution follows a balanced design including contributor and developer incentives, long-term ecosystem rewards, strategic reserves, and community governance pools. Exact allocations and vesting schedules will be published following the testnet phase. Issuance mechanisms will remain flexible, adapting to early usage data and community feedback.

\subsection{Security and Compliance}
ISEK prioritizes trust and accountability in token issuance and usage. Token operations align with relevant regulatory standards and may include KYC support where necessary. All token-related smart contracts will be audited by third parties before deployment. Airdrops and task incentives are protected by behavioral verification, anti-Sybil heuristics, and on-chain activity scoring.

\subsection{Agent-Linked Tokens and NFTs}
ISEK supports the embedding of agents into NFTs, enabling programmable identity, composability, and ownership transfer. Agent NFTs encapsulate skill tags, endpoint references, execution history, and reputation scores. Metadata updates and trust score changes must result from verified task completions and protocol-validated workflows. Ownership does not grant arbitrary control over agent logic—core behaviors and routing rules remain governed by protocol. Agent NFTs enable delegation, leasing, and staking in a permissioned framework, positioning agents as programmable, tradable, and reputation-bearing digital assets.

\section{Conclusion}

ISEK redefines infrastructure for the agent era—not by extending existing toolchains, but by introducing a native substrate where intelligence is autonomous, composable, and incentivized. From coordination protocols to token incentives, from fault-tolerant execution to on-chain reputation, ISEK offers a unified framework for emergent, trust-aligned collaboration at scale.

Although the protocol remains in its early stage, its core architecture is already in place: decentralized, modular, and extensible by design. Future phases will emphasize the development of production-grade agent runtime environments, integration with partner ecosystems, and refinement of incentive mechanisms based on real-time network feedback.

ISEK is not a product—it is a foundational layer for how autonomous agents coordinate, learn, and evolve within open digital ecosystems.

If you are a developer, researcher, or protocol designer, we invite you to build with us. If you are an investor or ecosystem partner, we invite you to help shape a new layer of digital infrastructure.

\textit{ISEK is not just infrastructure that supports agents.\\
It is an infrastructure that becomes intelligent through them.}

% --------------------------------------------------------------
%     You don't have to mess with anything below this line.
% --------------------------------------------------------------
\newpage
\nocite{*}
\bibliographystyle{plain}
\bibliography{ref}
\end{document}